# A Simple and Scalable Graphene Patterning Method and Its Application in CdSe Nanobelt/Graphene Schottky Junction Solar Cells

*Yu Ye [a], Lin Gan [b], Lun Dai [a,*], Yu Dai [a], Xuefeng Guo [b], Hu Meng [a], Bin Yu [a], Zujin Shi [b], and Guogang Qin [a,*]*

a State Key Lab for Mesoscopic Physics and School of Physics, Peking University, Beijing 100871, China

b College of Chemistry and Molecular Engineering, Peking University, Beijing 100871, China

\* Corresponding author. E-mail: lundai@pku.edu.cn; qingg@pku.edu.cn.

We have developed a simple and scalable graphene patterning method using electron-beam lithography (EBL) or ultraviolet (UV) lithography followed by a lift-off process. This method is free from additional etching or harsh process, universal to arbitrary substrates, compatible to Si microelectronic technology, and easy to be utilized to diverse graphene-based devices. Besides, it has the merits of high pattern resolution and alignment accuracy, which is especially useful for practical application, where manufacturing large-area graphene pattern to designated location with high resolution and alignment accuracy is highly desired.

Figure 1 shows a schematic illustration of the simple and scalable graphene patterning processes. First, large-scale graphene was synthesized on a Cu foil by the CVD method (Fig. 1a). After a layer of PMMA was spun on the graphene (Fig. 1b), the underlying Cu foil was etched using 1M $FeCl_3$ solution (Fig. 1c) for 1-2 h. The graphene/PMMA film floating in the etchant was then collected manually and placed into deionized (DI) water. The device substrate was prepared as follows: First, a layer of PMMA was spun on it (Fig. 1d). Then, EBL (FEI Strata DB 235) was employed to pattern the PMMA into desired shapes with high resolution at desired locations with high alignment accuracy (Fig. 1e). After that, the graphene/PMMA film floating on the DI water was manually collected on to the patterned device substrate (Fig. 1f). The flexibility of the graphene/PMMA film lets it conform into the windows area and contact with the bare device substrate. A heat treatment (90 °C, 30 min) in an oven was carried out for enhancing the contact between the graphene and substrate. Last, the PMMA together with the above graphene was removed by a lift-off process in acetone (Fig. 1g), and the graphene pattern was formed on the device substrate. It is worth noting that UV lithography, which has the advantage of manufacturing large-area graphene pattern on basically arbitrary substrates, can also be used instead of EBL in this method (as demonstrated below).

Figure 2a-e show field-emission scanning electron microscope (FESEM) images of the patterned-graphene fabricated using EBL with the shapes of micro-scale squares, triangles, line arrays, letters A, and letters B, respectively, on $Si/SiO_2$ substrates. The edges of these patterns are sharply demarcated from the substrates. Here, the minimum linewidth is ~ 2 μm. It is worth noting that in this work we have not optimized the process parameters to reach the linewidth limit yet. The right panel of figure 2f shows the optical image of a patterned-graphene fabricated with UV lithography on a flexible transparent substrate. We can see that the macro-scale graphene pattern faithfully copies the feature on the photomask (the left panel).

Before utilizing this graphene patterning method to fabricate CdSe NB/graphene Schottky junction solar cells, the synthesized graphenes are characterized. Figure 3a shows a typical Raman spectrum of the graphenes. It shows two main peaks: a narrow linewidth (~27 $cm^{-1}$) G-band peak (~1594 $cm^{-1}$) and a narrow linewidth (~45 $cm^{-1}$) 2D-band peak (~2689 $cm^{-1}$). Their intensity ratio ($I_{2D} : I_G$) is about 2.4, indicating the formation of monolayer graphene [1]. Besides, the defect-related [2] D-band peak (~1358 $cm^{-1}$) is weak, indicating the high quality of the graphene. Figure 3b shows the transparency spectrum of the graphene. It can be seen that high transparency (>98%) occurs in the range of wavelength longer than 500 nm. The typical sheet resistance of the graphenes is about 345 Ω/□.

CdSe, due to its suitable bandgap (1.74 eV) for solar light absorption, is an important photovoltaic (PV) material, and has attracted more and more attention recently [3, 4]. We have utilized the developed graphene patterning method to fabricate CdSe nanobelt (NB)/graphene Schottky junction solar cells. Figure 4a shows the schematic illustration of the CdSe NB/graphene Schottky junction solar cell. One In/Au (10/100 nm) ohmic electrode was defined at one end of the CdSe NB by UV lithography, thermal evaporation, and lift-off processes. The graphene film was made on the other end of the NB with the developed graphene patterning method using UV lithography. A FESEM image of an as-fabricated CdSe NB/graphene Schottky junction solar cell is shown in Fig. 4b. The Au (100 nm) electrode, which forms an ohmic contact with the graphene, was used for later-on welding purpose.

The mechanism of Schottky junction solar cell can be understood qualitatively by plotting the energy band diagram. Figure 5a shows the energy diagram of a CdSe NB/graphene Schottky junction solar cell under illumination. Due to the work function difference between the graphene ($\Phi_G$ ~4.66 eV) [5] and CdSe ($\Phi_{CdSe}$ ~4.2 eV) [6], a built-in potential ($eV_i$) forms in the CdSe near the CdSe NB/graphene interface. Under illumination, the photogenerated holes ($h^+$) and electrons ($e^-$) are driven toward the Schottky electrode (graphene film) and CdSe NB, respectively, by the built-in electric field. When the solar cell is open-circuited, the separated photogenerated electrons and holes will produce an open-circuit voltage $V_{OC}$. When the solar cell is short-circuited, the extracted photogenerated carriers will transit through the external circuit, generating a short-circuited current $I_{SC}$.

Figure 5b shows the room-temperature *I-V* characteristic (in dark) of the solar cell depicted in Fig. 4b on a semi-log scale. We can see that the *I-V* curve shows an excellent rectification characteristic. An on/off current ratio of ~ $7.5\times10^4$ can be obtained when the voltage changes from +1 to −1 V. For Schottky junction diodes made on high-mobility semiconductors, such as Si and CdSe etc., the current $I$ is determined by the thermionic emission of electrons and can be described by the equation $I = I_0[\exp(eV/nkT)-1]$ [7], where $I_0$ is the reverse saturation current, $e$ is the electronic charge, $V$ is the applied bias, $n$ is the diode ideality factor, $k$ is the Boltzmann's constant, $T$ is the absolute temperature. By fitting the measured $I$-$V$ curve with the above equation, we obtain $n$=1.17. Here, $n$ is quite close to the value of an ideal Schottky junction ($n$=1) [8]. All these results indicate that the graphene film has formed a good Schottky contact with the CdSe NB.

Figure 5c shows the room-temperature $I$-$V$ characteristic of the solar cell depicted in Fig. 4b under AM 1.5G illumination. We can see that this cell exhibits an excellent PV behavior with open-circuit voltage ($V_{OC}$) of about 0.51 V, short-circuit current density ($J_{SC}$) of about 5.75 mA/cm$^2$, fill factor ($FF$) of about 42.7%. The current density is calculated based on the effective Schottky junction area (the overlapping area of CdSe NB and graphene: ~1988.6 μm$^2$), where the charge separation takes place [6]. Therefore, we can estimate the corresponding energy conversion efficiency ($\eta=FF\cdot J_{SC}\cdot V_{OC}/P_{in}$, $P_{in}$: illumination light power density, 100 mW/cm$^2$) to be about 1.25%. We attribute the high performance of the CdSe NB/graphene Schottky solar cell to the high-performance graphene Schottky electrode fabricated with the developed graphene pattering method.

We have developed a simple and scalable graphene patterning method using electron-beam or UV lithography followed by a lift-off process. This method will accelerate the application of graphene in diverse optoelectronic and electronic devices, especially in large-scale graphene array-based applications. Novel CdSe NB/graphene

Schottky junction solar cells were fabricated using this developed graphene patterning method. Typical as-fabricated solar cells show excellent PV behavior with the AM 1.5G energy conversion efficiency up to 1.25%. These high-performance CdSe NB/graphene Schottky junction solar cells are promising for novel third generation solar cells, which have potential application in integrated nano-optoelectronic systems.

**Acknowledgement**

This work was supported by the National Natural Science Foundation of China (Nos.10774007, 11074006, 10874011, 50732001), the National Basic Research Program of China (Nos.2006CB921607, 2007CB613402), and the Fundamental Research Funds for the Central Universities.

**References**

[1] Reina, A.; Jia, X. T.; Ho, J.; Nezich, D.; Son, H.; Bulovic, V.; Dresselhaus, M. S.; Kong, J. *Nano Lett.* **2009**, 9, 30-35.

[2] Ferrari, A. C. *Solid State Commun.* **2007**, 143, 47-57.

[3] Yu, Y. H.; Kamat, P. V.; Kuno, M. *Adv. Funct. Mater.* **2010**, 20, 1464-1472.

[4] Lee, Y. L.; Lo, Y. S. *Adv. Funct. Mater.* **2009**, 19, 604-609.

[5] Shi, Y. M.; Kim, K. K.; Reina, A.; Hofmann, M.; Li, L. J.; Kong, J. *ACS Nano* **2010**, 4, 2689-2694.

[6] Zhang, L, H.; Jia, Y.; Wang, S. S.; Li, Z.; Ji, C. Y.; Wei, J. Q.; Zhu, H. W.; Wang, K. L.; Wu, D. H.; Shi, E. Z.; Fang, Y.; Cao, A. Y. *Nano Lett.* **2010**, 10, 3583-3589.

[7] Sharma, B. L. *Metal-Semiconductor Schottky Barrier Junctions and Their Applications*; Plenum Press: New York, 1984.

[8] Sze, S. M. *Semiconductor Devices: Physics and Technology, 2nd ed.*; John Wiley & Sons: New York, 2001.

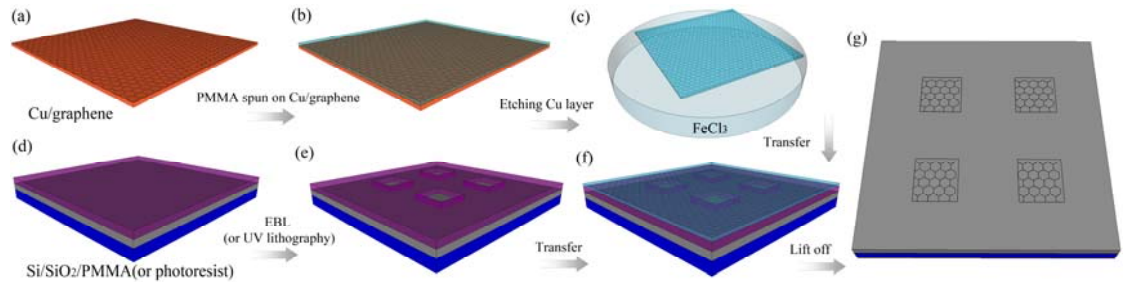

Figure 1. Schematic illustration of the simple and scalable graphene patterning processes. (a) Synthesizing a large-scale graphene film on a Cu foil. (b) A layer of PMMA was spun on the Cu foil with the synthesized graphene on the top. (c) The underlying Cu film was etched using 1M $FeCl_3$ solution. (d) A layer of PMMA (or photoresist) was spun on a device substrate. (e) EBL (or UV lithography) was employed to pattern the PMMA (or photoresist) into desired shapes at desired locations. (f) The graphene/PMMA film was manually collected onto the patterned device substrate. (g) PMMA (or photoresist) together with the above graphene was removed by a lift-off process in acetone, and the patterned graphene was formed on the device substrate.

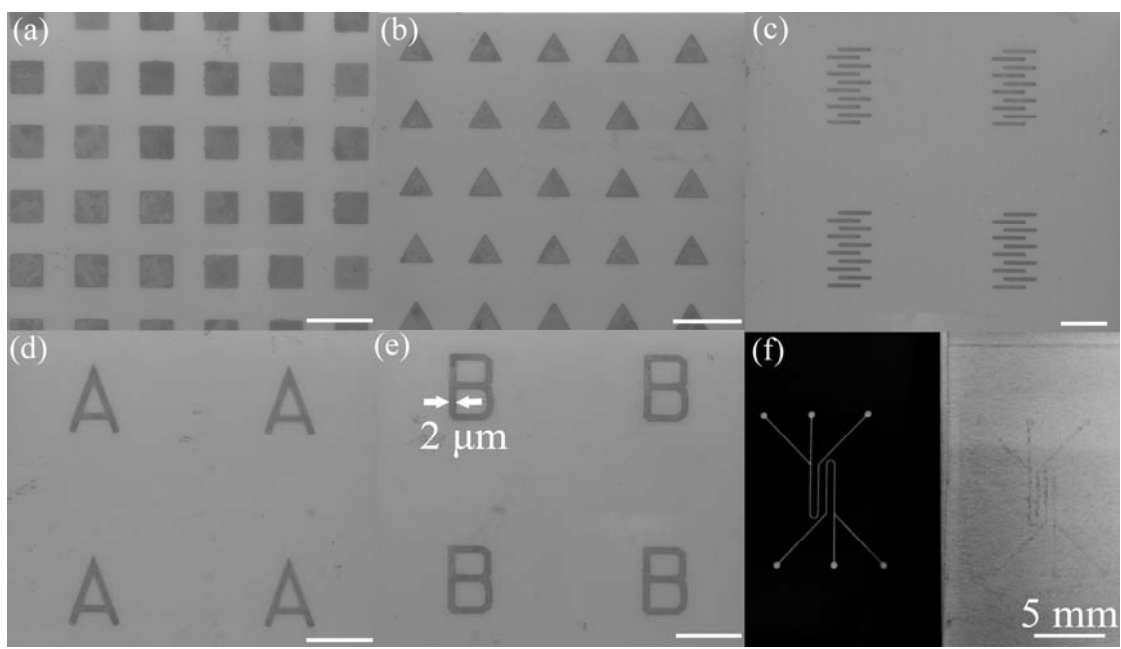

Figure 2. (a)-(e) FESEM images of the patterned graphene by EBL with shapes of micro-scale squares, triangles, line arrays, letter A, letter B, respectively, on SiO$_2$/Si substrates. Scale bars: 20 μm. (f) Left panel: optical image of the photomask. Right panel: optical image of the patterned graphene by UV lithography on a flexible transparent substrate.

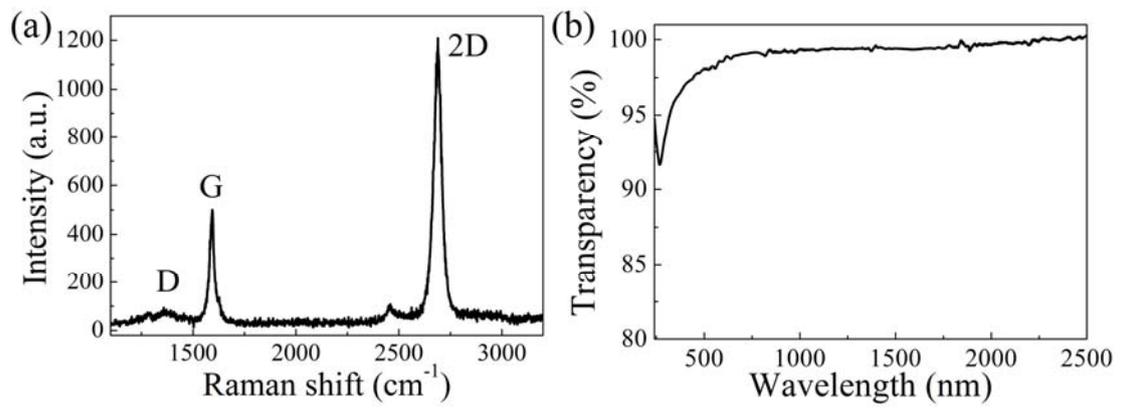

Figure 3. (a) Raman Spectrum of the graphene on a Si/SiO$_2$ substrate. (b) Transparency spectrum of the graphene on a quartz substrate.

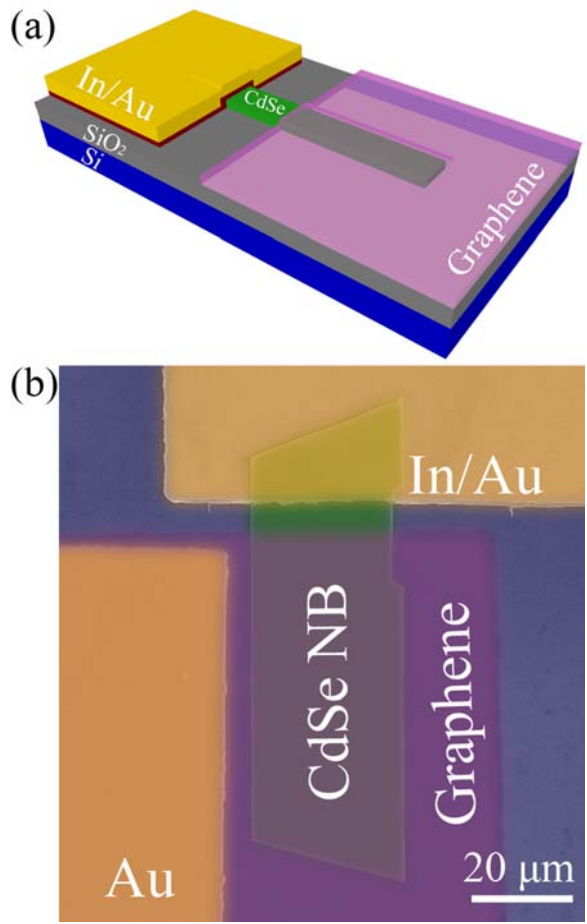

Figure 4. (a) Schematic illustration of the CdSe NB/graphene Schottky junction solar cell. (b) FESEM image of a CdSe NB/graphene Schottky junction solar cell.

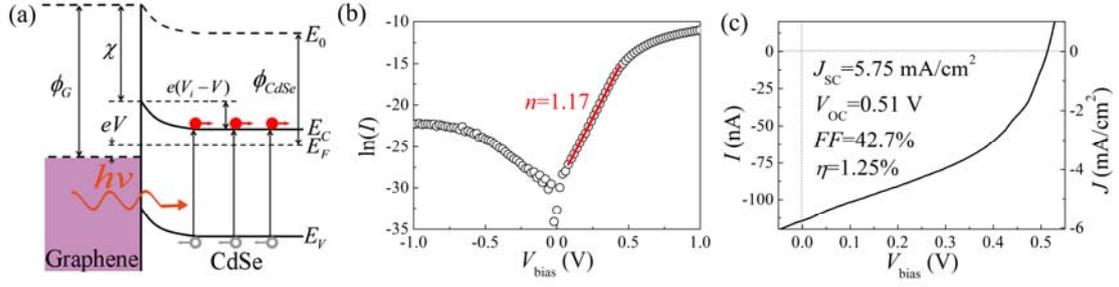

Figure 5. (a) Energy diagram of a CdSe NB/graphene Schottky junction solar cell under illumination. $\Phi_G$, $\Phi_{CdSe}$ are the work functions of graphene and CdSe, respectively. $eV_i$ is the built-in potential. $V$ is the output voltage of the solar cell. $\chi$ is the electron affinity of CdSe. $E_C$, $E_V$, $E_F$ correspond to the conduction band edge, valence band edge, and Fermi level of CdSe, respectively. $E_0$ corresponds to the vaccum level. (b) Room-temperature *I-V* characteristic (in dark) of the CdSe NB/graphene Schottky junction solar cell depicted in Fig. 4b on a semi-log scale. The red straight line shows the fitting result of the *I-V* curve by the equation $\ln(I) = eV/nkT + \ln(I_0)$. (c) Room-temperature *I-V* characteristic of the same solar cell under the AM 1.5G illumination with light intensity of 100 mW·cm$^{-2}$.